\documentclass{iopart}
\usepackage{iopams}
\usepackage{setstack}
\usepackage{graphicx}
\usepackage{epsfig}
\begin{document}
\title[Proposed method for searches of gravitational waves  from blazars]{Proposed method for searches of gravitational waves   from PKS 2155-304 and other blazar flares}
\author{S~Desai$^{1}$, K~Hayama$^{2}$, S~D~Mohanty$^{2}$, M~Rakhmanov$^{2}$, T~Summerscales$^{3}$, S Yoshida$^{4}$}

\address{
$^{1}$ The Pennsylvania State University, University Park, PA 16802, USA}
\address{
$^{2}$ The University of Texas, Brownsville, TX 78520, USA}
\address{
$^{3}$ Andrews University, Berrien Springs, MI 49104, USA}
\address{
$^{4}$ Southeastern Louisiana University, Hammond, LA 70402, USA}

\ead{desai@gravity.psu.edu} 
\begin{abstract}

We propose to  search for gravitational waves 
from PKS 2155-304 as well as other blazars. 
PKS 2155-304  emitted a long duration energetic flare in July 2006, 
with   total isotropic equivalent energy released in TeV
gamma rays  of approximately $10^{45}$ ergs. Any possible gravitational wave 
signals associated 
with this outburst should be seen by  gravitational wave 
detectors at the same time as the electromagnetic signal. 
During  this flare, the two LIGO interferometers at 
Hanford  and the GEO detector were in operation  and collecting data. For this
search  we will  use the data from multiple gravitational wave detectors. 
The method we use for this purpose is a coherent network analysis algorithm and is called {\tt RIDGE}. To estimate the sensitivity of the search, we perform numerical simulations. The 
sensitivity to estimated gravitational wave energy at the source is about $2.5 \times 10^{55}$ ergs for a 
detection probability of 20\%.  For this search,  an end-to-end analysis 
pipeline has been developed, which takes into account the motion of the 
source across the sky.

\end{abstract}

\section {Introduction}

Blazars are one  type  of active galactic nuclei which  are powered by accretion
onto a central engine,  and have been detected throughout the electromagnetic spectrum
from radio waves to TeV gamma rays~\cite{urry}.
Blazars show strong variability on many different time scales throughout
the electromagnetic spectrum and exhibit a high degree of polarization. Their
 radio jets exhibit apparent super-luminal motion indicating that they  are
emitted at small angles to our line of sight. 
The central engine is believed to consist of   a super-massive 
black hole ($>10^6 M_{\odot}$)
or a  binary black hole~\cite{Hayasaki}. There are more 
than 500 confirmed blazars and their mean redshift is about 0.2~\cite{Turiziani}.

In many respects blazars are similar to 
gamma-ray bursts (GRBs), and there have been many searches for gravitational waves
from GRBs~\cite{s2s3s4}. Both  GRBs and blazars have a central engine and a jet, both 
undergo accretion,  and both show evidence for non-thermal emission~\cite{Ghishellini06}.
The main difference between them  is the nature of  the central engine, and that GRBs emit outbursts only once, whereas blazars emit outbursts more than once.

Spatial correlations have been observed between blazars and 
ultra high energy cosmic rays~\cite{Gorbunov}.
There have been searches for  high energy neutrinos associated with   
blazar flares~\cite{Abe}. The benefits  of coincidence searches between the data from neutrino detectors and gravitational wave detectors  are outlined in ~\cite{Aso}. This also motivates the search for gravitational waves in coincidence 
with blazar flares. 

In this paper we present a  method for searching for gravitational waves  
associated with an energetic outburst from PKS 2155-304 blazar, which happened 
during the fifth LIGO science run (S5)~\cite{LIGO}. Assuming that the velocity of 
gravitational waves is same as that of light, any possible gravitational 
wave signals should be seen simultaneously along with the electromagnetic outburst. Therefore, this search needs to be done by analyzing the data from gravitational wave detectors at the time of this flare.

To do this search, we shall combine the data from many gravitational 
wave detectors. The search will be done with a coherent network analysis algorithm  called {\tt RIDGE}~\cite{Hayama07}.
Sensitivity results using simulated noise and one set of simulated 
signals are presented. We also briefly discuss the outbursts 
seen from OJ 287 and S5 0716+71, which are two other blazars we plan to search for gravitational 
waves. 

Since the flare from PKS 2155-304 lasted several hours, the search for gravitational wave signals 
from this object has features  that are quite distinct from  previous LIGO 
searches for gravitational waves associated with GRBs.  
For this search, we need to analyze a much longer data segment,  during which the source moves significantly on the sky.
This is different from  searches for gravitational waves from GRBs, where 
the on-source time interval   for which data is analyzed for possible gravitational wave signals is only  180 seconds~\cite{s2s3s4}. 
For this search, the corresponding change in the detector responses due to the changing 
antenna pattern functions must be taken into account.

\section {PKS 2155-304 outburst in July 2006}
\label{sec:pks}
PKS 2155-304 is a blazar located in the southern galactic hemisphere with redshift of about 0.12, which corresponds to a luminosity distance of about 540 Mpc~\cite{Chadwick}, and  has been observed throughout the electromagnetic spectrum. 
The nature of the central engine in PKS 2155-304 is currently unknown.
The High  Energy Stereoscopic System (H.E.S.S)~\cite{HESS} detector has been monitoring
PKS 2155-304 since 2002. It 
detected an energetic outburst from PKS 2155-304~\cite{Aharonian07} above 200 GeV, which lasted about 2000
seconds, starting from MJD = 53944 (corresponding to July 28 2006 00:40:00 UTC). The total flux during this flare increased by a factor of ten with respect to the quiescent level. The integrated flux during  this flare is $   1.7 \times 10^{-9} \mbox{cm}^{-2} \mbox{sec}^{-1}$. 
The total isotropic equivalent energy emitted during this flare assuming 
the source has a redshift of about 0.12, matter density of  0.3, density of cosmological constant of 0.7, is 
approximately $10^{45}$  ergs.

Various models have been proposed for this outburst~\cite{Rees,Ghishellini,Dermer}, but none of them could 
satisfactorily explain this large flare along with the steady state emission from this object.
The common feature of all the models used to explain the flare is that the bulk Lorentz factor of 
the jet is around 50, size of the emitting region is around $3 \times 10^{14}$ cm,
and  the jet opening angle is about $1^{\circ}$, which reduces the isotropic equivalent 
energy emitted by a factor of about  $10^{-4}$~\cite{Kulkarni99}. 

\begin{figure}
\begin{center}
\includegraphics[width=0.5\textwidth]{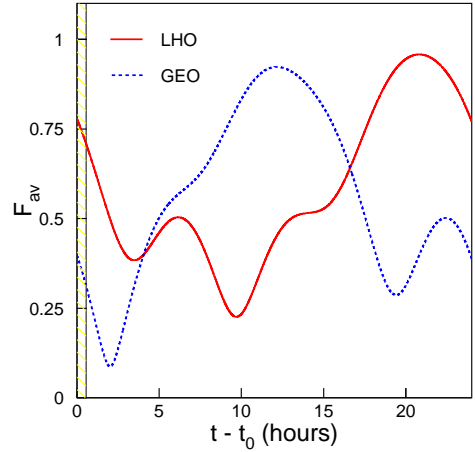}
\caption{The detector response function ($F_{av} = \sqrt{F_{+}^2 + F_{\times}^2}$) of the LIGO Hanford detectors and GEO 
detector in a 24 hour interval starting from the PKS 2155-304 flare, which corresponds to GPS time ($t_0$) equal to  838082414.
The hatched region shows the duration of the flare}
\label{antresp}
\end{center}
\end{figure}

During this outburst from PKS 2155-304, both the 4~km
(H1) and 2~km (H2) LIGO Hanford interferometers and the GEO detector were operational. The 
third LIGO detector in Livingston, LA 
was not taking data at this time. The angular sensitivities to unpolarized gravitational waves of H1, H2, and GEO within a one day interval from the start of this flare, are shown in 
figure~\ref{antresp}. The sensitivity of gravitational wave interferometers can be characterized by its
{\it Inspiral Range}~\cite{Sutton}, defined as the distance to which  
signals from the inspiral of two $1.4 M_{\odot}$ neutron stars is detected with signal-to-noise ratio greater than 8, after 
averaging over all sky positions and orientations of the binary system. The
corresponding  Inspiral Range for H1, H2, and GEO detectors were 13 Mpc, 7 Mpc and
1.3 Mpc respectively during this time. This is at least 50 times smaller than the 
distance to PKS 2155-304.

\section{Mechanisms for possible gravitational wave  emission}
\label{mechanism}
There are  no theoretically accepted gravitational wave emission 
mechanisms from such transient blazar
outbursts in the LIGO frequency band. There have been some proposed mechanisms for steady state
gravitational wave emission from Active Galactic Nuclei with super-massive black holes as the central engine  in the LISA frequency band, which we briefly describe here. We should point out  that the distance to this object 
is much greater than the Inspiral Range of the LIGO and GEO detectors (Section~\ref{sec:pks}) 
at this time,  and a large amount of gravitational wave energy must be emitted if the 
expected signal is near the sensitivity of initial LIGO. Ultimately, since the cause of this 
flare is unknown, we would like to be open to all possibilities and look for gravitational 
waves during this outburst without any theoretical prejudice.

One mechanism is through the fragmentation of the accretion disk  due to  feedback energy from star formation in the outer parts of the accretion disk for a blazar with single black hole~\cite{Levin06} or  due to impact from the accretion disk of the secondary black hole for a blazar with binary black hole~\cite{Valtonen06}. 
 Another mechanism is through the fragmentation of the accretion disk because of 
dynamical friction from an in-falling satellite onto a coplanar accretion disk~\cite{Chang08}.
This fragmentation results  in the formation and evolution of massive 
stars in the self-gravitating accretion disks of massive black 
holes. The resultant compact objects from the fragmentation  remain embedded 
in the accretion disk and could merge with the parent black hole at the center~\cite{Levin06}.
The gravitational waves are possibly produced  from inspirals and mergers of the disk 
born compact objects (with masses of about 100~$M_{\odot}$) with the central black hole. 
The expected strain amplitude for the above mechanisms has been
estimated to be from $10^{-21}$ to $10^{-19} \mathrm{Hz}^{-1/2}$ with expected 
frequency from $10^{-4}$ to $10^{-1}$ Hz~\cite{Sigl06}.
If a less massive object is formed, the expected frequency of the gravitational wave signal would be higher 
and could fall into the LIGO band.
However, a less massive object will also lead to a significant reduction in the amount of 
gravitational wave emission.



\section{Analysis  techniques}
Since the LIGO Hanford  and GEO detectors are far apart, we will 
use gravitational wave coherent network analysis techniques for 
analyzing this flare. Such  coherent network analysis algorithms take 
into account the antenna patterns for each detector in order to extract
the gravitational wave signal from the detector data stream~\cite{Rakhmanov05,x,cwb,flare}.
The {\tt RIDGE}~\cite{Hayama07} algorithm (the name is 
derived from the term ``ridge regression'' used in 
statistics literature) is one such method. The technique 
solves the problem of rank deficiency of the antenna response matrix by  
using Tikhonov regularization~\cite{Rakhmanov05}.  More details on   
{\tt RIDGE}  and  the data conditioning technique used  can be found in~\cite{Hayama}. We also propose to 
use the same method in searches for gravitational waves from GRBs, 
magnetars, pulsar glitches, and Sco-X1.  
For this analysis we would like to combine the data from the GEO 
detector with the data from the two Hanford detectors. The GEO detector is 
geographically far apart  and has different arm orientations with  respect to  
the Hanford detectors. Therefore,  using GEO data would enable us to increase the sensitivity 
in contrast  to searches for gravitational waves   from GRB 070201~\cite{070201} or SGR 1806-20~\cite{sgr1806}, which 
used the data from only the Hanford detectors.

We are also developing a method to look for long-duration (lasting several 
seconds to minutes) transient gravitational wave signals~\cite{Hayama08}, taking into 
account the motion of the source in the sky. This method of source tracking involves splitting the dataset into many 
short duration segments during which the position of the source is approximately constant. We 
reconstruct $h_{+}$ and $h_{\times}$ within each segment knowing the  direction to the source. 
The reconstructed $h_{+}$ and $h_{\times}$ is then concatenated  across the segments and a test for excess power 
is applied to  this reconstructed time-series.
 Since the flare from PKS 2155-304 lasted for more than 
 two hours, we plan to  apply this source tracking technique.  
This method of looking for long duration unmodelled bursts in LIGO data  could help 
us find a signal which otherwise may be missed from all sky untriggered searches for 
gravitational wave bursts~\cite{s4burst}, which will also analyze the data during this flare.

\section{Estimation of sensitivity of the algorithm}
In order to evaluate the performance of this algorithm, we perform numerical
simulations of the detector noise and possible gravitational wave signals.
We then apply the {\tt RIDGE} algorithm on this simulated time-series 
which includes both the detector noise and signal. We shall then estimate the sensitivity of our search.
using Receiver Operating Characteristic curves.

Since the flare from PKS 2155-304 lasted 2000 seconds, we need to generate 
the same duration of simulated noise for H1, H2, and GEO.  The 
amplitude spectral density of the simulated noise time series is obtained 
from the LIGO and GEO design curves. 

\begin{figure}
\begin{center}
\includegraphics[scale=0.5]{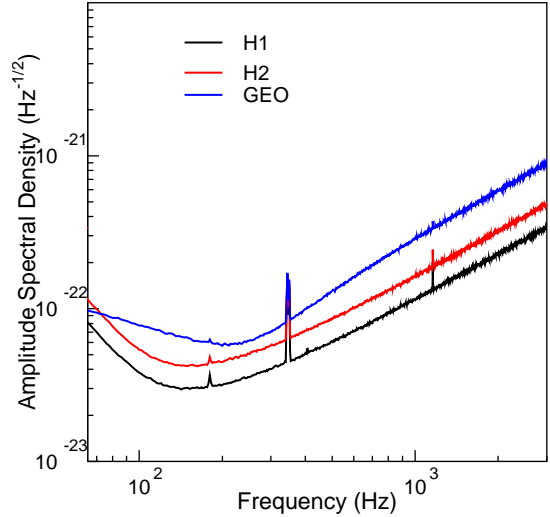}
\caption{The amplitude spectral density of simulated data for the 4 km and 2 km Hanford interferometers (H1 and H2)  and 
GEO  detector (assuming signal detuning frequency of 200 Hz) as a function of frequency.}
\label{psd}
\end{center}
\end{figure}

We first generate 2000 seconds of  Gaussian stationary noise 
using three independent realizations of white noise corresponding to the 
three detectors.  We then created FIR filters with transfer functions 
matching the design amplitude noise spectral density curves as a function of frequency,
 for the three instruments.
These were applied to the generated stationary Gaussian noise.
To generate a realistic simulation of the detector noise we also need to add
possible instrumental lines. To model the effect of such lines, sinusoidal signals were added 
at seven frequencies between 50 and  1050 Hz. The 
amplitude spectral density of the simulated noise for LHO and GEO is shown 
in figure~\ref{psd}. 

In order to test the performance of this method to possible short-duration gravitational wave signals
from PKS 2155-304, we need to add simulated burst signals to the generated noise 
at various times throughout the flare. In its simplest form, a burst signal can be described by a sine-Gaussian 
waveform, with quality factor $Q$, central frequency ($f_0$) and a characteristic 
amplitude~\cite{LIGOs4}. The waveforms we chose for our simulation  were circularly polarized sine-Gaussian
signals with $Q = 9$ and $f_0 = 235$~Hz, and for which the signal in `$+$' polarization is phase shifted by $90^{\circ}$ 
with respect to the signal in `$\times$' polarization. 
The strength of the signal is characterized by
\begin{equation}
h_\mathrm{rss}^2 = \int_{-\infty}^{\infty} \rmd t\; \left(h_+^2(t) + 
h_\times^2(t)\right)  ,
\end{equation}
where $h_{+}$ and $h_{\times}$ correspond to waveforms with the `+' and `$\times$' polarizations 
respectively.
We chose four different values for the    signal strength with  $h_\mathrm{rss}$ equal to 
(0.7, 1.4, 2.1, 3.5) $\times 10^{-22} {\rm Hz}^{-1/2}$,
This signal was injected at the estimated location of PKS 2155-304,
which is Right Ascension  = 22 hour and  Declination = $-29.8^{\circ}$~\cite{Chadwick}.  

The {\tt RIDGE} method is now applied to our simulated dataset.
 At each  value of latitude ($\theta$) and longitude ($\phi$), we maximize the 
Tikhonov regularized likelihood functional~\cite{Rakhmanov05}, by varying $h_{\times}$ and
$h_{+}$. To distinguish between signals and noise, we define a detection statistic 
using the values of the likelihood functional (${\bf S}$), and is  called 
{\it radial distance statistic}~\cite{Hayama07}
\begin{equation}
R_{\rm rad} =  \left[ \left(\frac{\max_{\theta,\phi} {\bf S}}{\max_{\theta,\phi} \overline{\bf S_0}} -1 \right)^2 + \left(\frac{\max_{\theta,\phi} {\bf S}}{\min_{\theta,\phi} {\bf S}} \times \frac{\max_{\theta,\phi} \overline{\bf S_0}}{\min_{\theta,\phi} \overline{\bf S_0}} -1 \right)^2\right]^{1/2},
\label{rad}
\end{equation}
where $\overline{\bf S_0}$ is the average likelihood of all skymaps that do not contain a signal.

\begin{figure}
\includegraphics[width=0.5\textwidth]{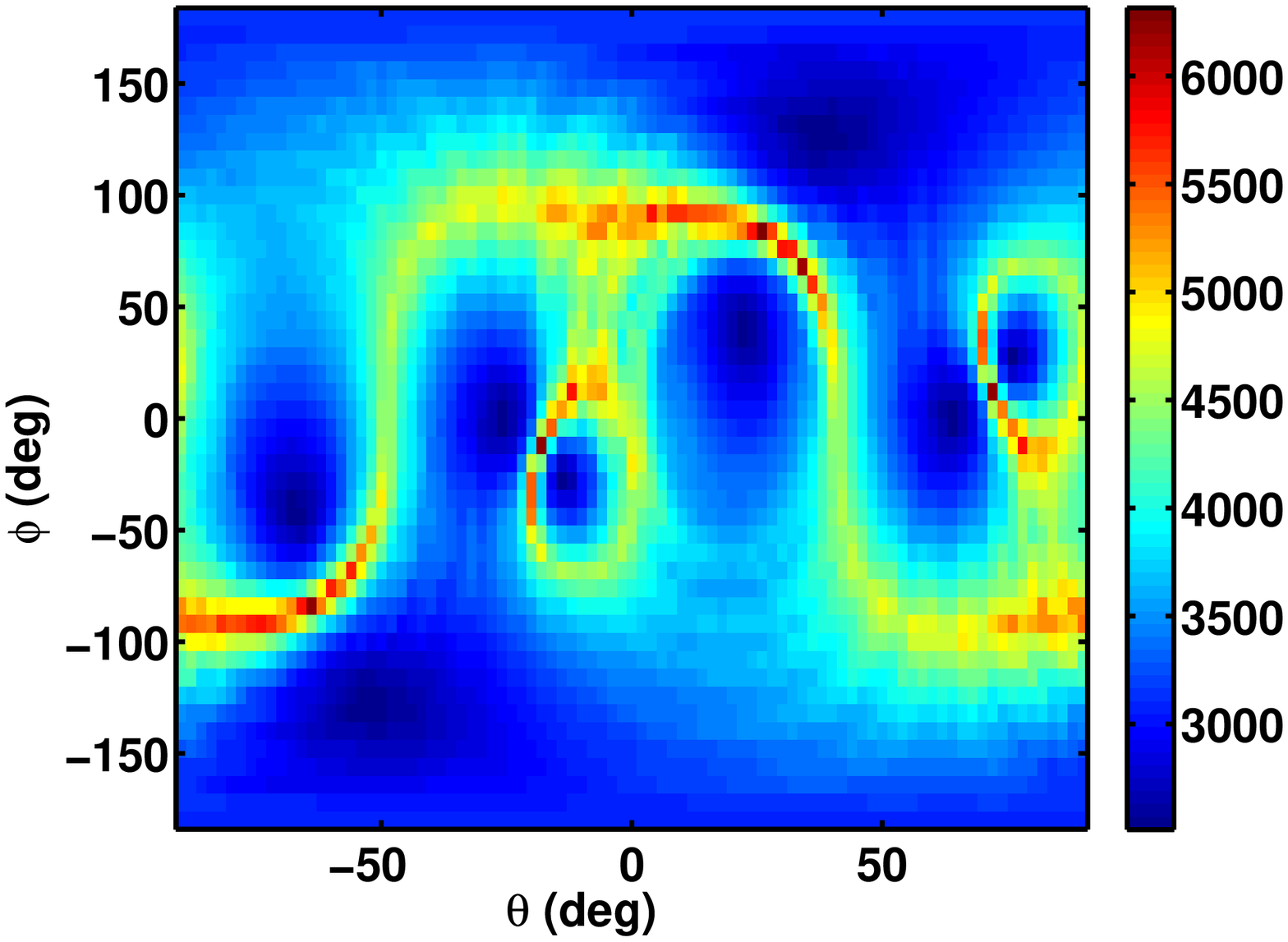}
\includegraphics[width=0.5\textwidth]{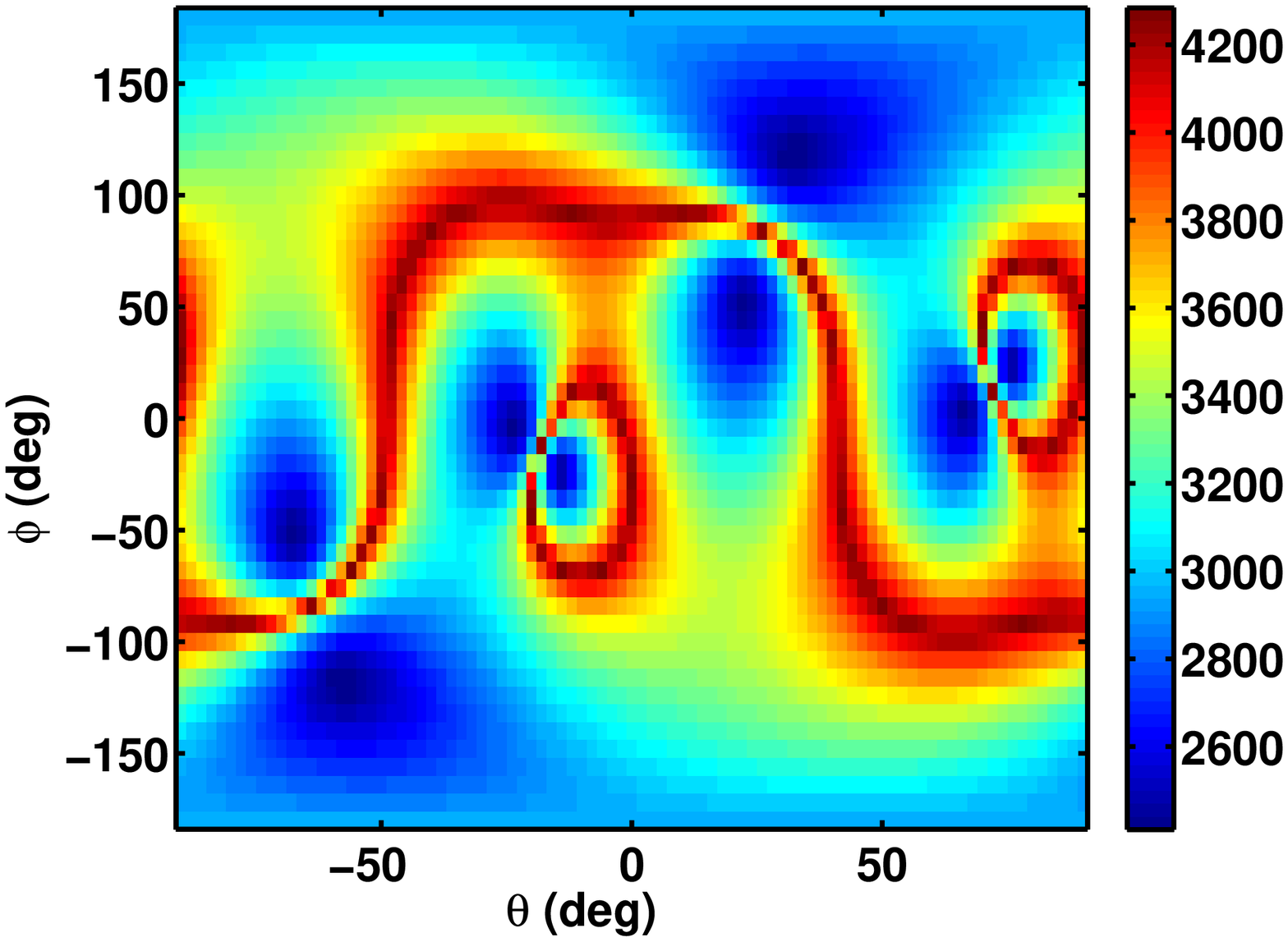} 
\caption{Likelihood skymap for a signal with $h_\mathrm{rss}=~3.5 \times 10^{-22} Hz^{-1/2}$ (left) 
and for noise only values (right) as a function of latitude($\theta$) and longitude ($\phi$).}
\label{skymap}
\end{figure}

\begin{figure}
\begin{center}
\includegraphics[width=0.5\textwidth]{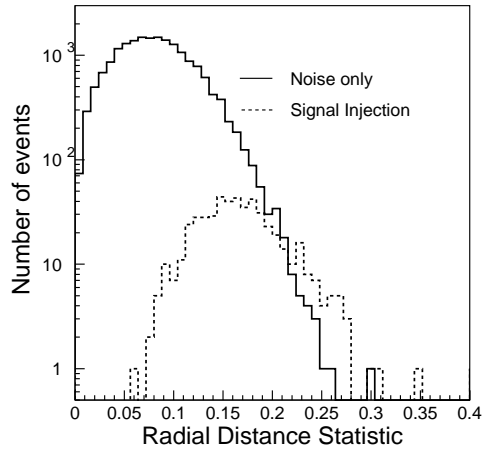} 
\caption{The distributions of the  radial distance statistic~(\ref{rad})  for both simulated  signals (dashed line) and noise (solid line). The strength of the injected signal is $h_\mathrm{rss}~=~2.1 \times 10^{-22} {\rm Hz}^{-1/2}$. The mean value of the radial distance statistic for signal injections is 
higher than  for pure noise.}
\label{radstat}
\end{center}
\end{figure}

In figure~\ref{skymap}, we show   plots of two skymaps in which the colour 
scale indicates the values of the likelihood functional.  
The first plot is  the skymap at the maximum of  the radial distance 
statistic which contains a signal. The second plot is  the skymap averaged over all times 
not containing any signal.  The pattern 
in the skymap is largely determined by the condition number of the LHO-GEO 
response matrix~\cite{Rakhmanov05}. 
The values of the radial distance statistic   for the signal only and noise-only skymaps are shown in 
figure~\ref{radstat}. 

\begin{figure}
\begin{center}
\includegraphics[width=0.5\textwidth]{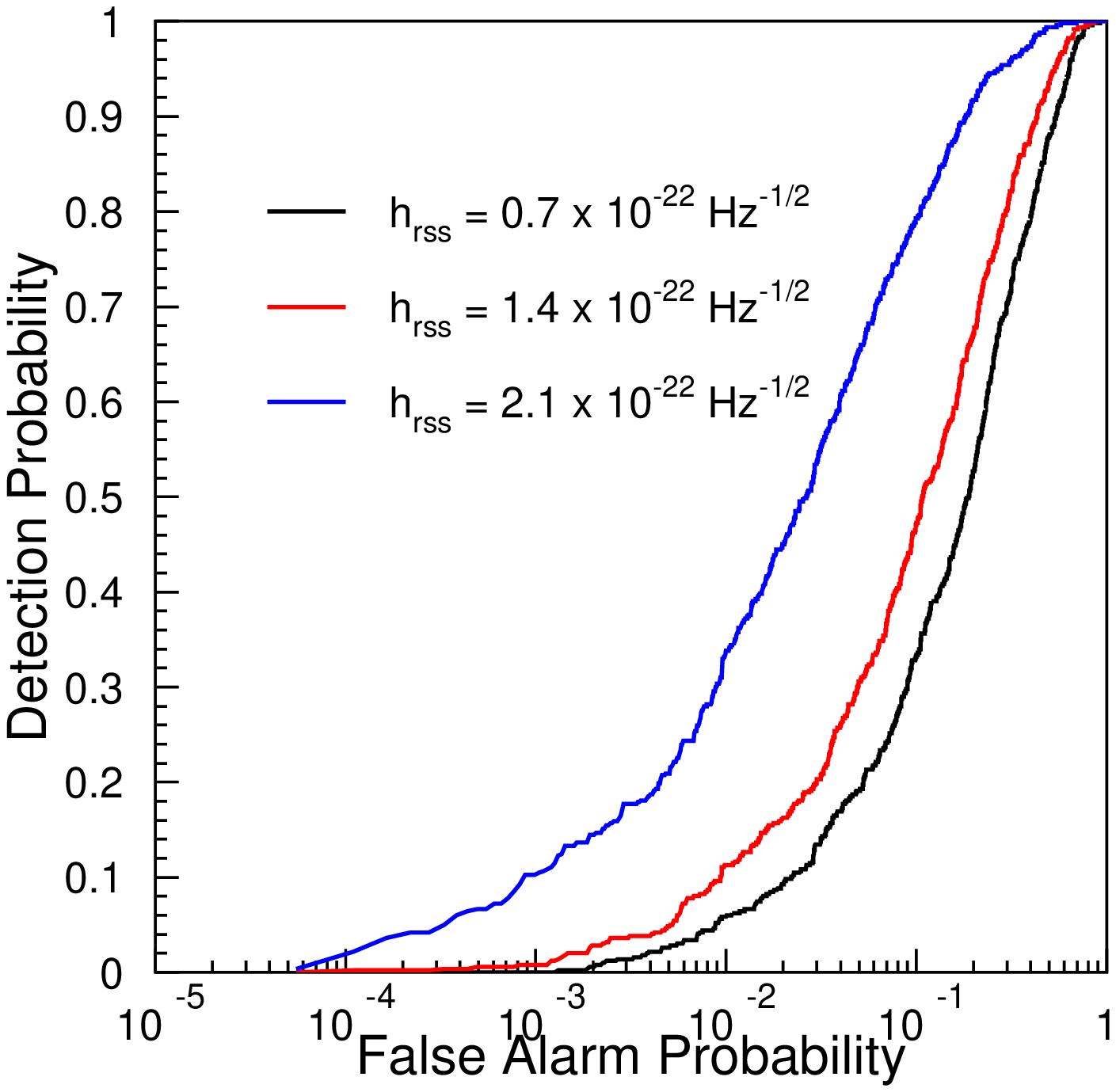} 
\caption{Receiver Operating Characteristic (ROC) curve for H1-H2-GEO network with the {\tt RIDGE} 
pipeline 
during the PKS 2155-304 flare for a circularly polarized sine-Gaussian signal for three different
values of $h_\mathrm{rss}$ shown in the figure in units of ${\rm Hz}^{-1/2}$.}
\label{roc}
\end{center}
\end{figure}

A common way to characterize the sensitivity of an algorithm is through Receiver Operating
Characteristic (ROC) curves of detection efficiency versus false alarm 
probability. The ROC curves of this search  with the  H1-H2-GEO network 
for three values of $h_\mathrm{rss}$ are shown in figure~\ref{roc}. These ROC curves are obtained
by calculating the detection probability and false alarm probability for different threshold values of the 
radial distance statistic.  For a detection probability of 0.2
and false alarm probability of 0.004, the  $h_\mathrm{rss}$ is about $2.1 \times 10^{-22} {\rm Hz}^{-1/2}$. 
The energy for  a sine-Gaussian signal with central frequency ($f_0$) and $Q \gg 1$ is given by~\cite{LIGOs4}
\begin{equation}
E_{GW} = \frac{r^2 c^3}{4 G}  (2 \pi f_0)^2 h_\mathrm{rss}^2 ,
\label{egw}
\end{equation}
where $r$ is the distance to the source. From (\ref{egw}), the $h_\mathrm{rss}$  
sensitivity corresponds to total energy in gravitational waves at the source of 
approximately $2.5 \times 10^{55}$ ergs.  Thus, the total energy emitted
in gravitational waves from this object must be about $10^{10}$ 
greater than the emitted electromagnetic energy, in order to be sensitive to our method.

\section{Outbursts from OJ 287 and S5 0716+71}
In addition to PKS 2155-304, there were other blazar outbursts during S5, two of which we will
describe here.   We shall also look for possible gravitational waves from these two  blazars.

\begin{figure}
\begin{center}
\includegraphics[width=0.5\textwidth]{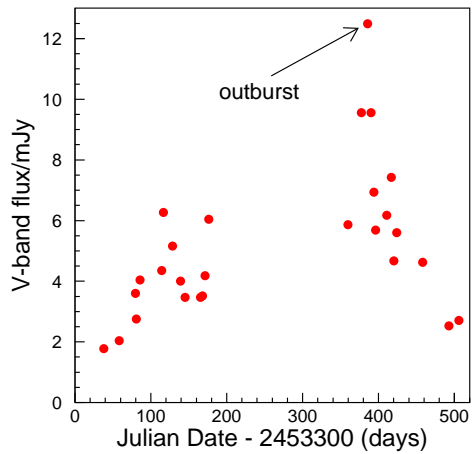} 
\caption{Observations from OJ 287 starting from 21 October 2004 12:00 UTC. The outburst is 
in November 2005 and corresponds to $V$-band flux of about 12 mJy. 
(Data used for this plot is obtained from ~\cite{Valtonen06}.)}
\label{oj287}
\end{center}
\end{figure}

OJ 287 is a blazar located at a redshift of  about 0.3, equivalent to  luminosity distance 
of about 1600 Mpc. 
OJ 287 has been known to emit periodic outbursts every 12 years. This blazar is 
believed to contain a binary black hole system with  masses equal to 
$16 \times 10^9 M_{\odot}$ and $0.1 \times 10^9 M_{\odot}$~\cite{Valtonen06}.
The possible cause of these bursts is due to the fragmentation of the accretion disk of the secondary black hole by the 
primary one. The orbital decay of the system due to the emission of gravitational radiation 
(with frequency approximately equal to $10^{-8}$ Hz) is in agreement with general relativity to within 10~\% accuracy~\cite{Valtonen08}. 
There were two optical outbursts from OJ 287 during S5, in November 2005 and September 2007.  
Figure~\ref{oj287}  shows one of  the bursts from OJ 287 in November 2005 in the optical $V$ band. 

Another blazar for which outbursts were observed in the radio, optical and gamma-ray bands is 
S5 0716+71. Not much is known about the redshift of this system or the nature of its central engine.
These outbursts were detected from August-September 2007 by the Whole Earth Blazar
Telescope Consortium using data from the AGILE satellite in MeV gamma rays, optical observations
in $R$-band, and radio observations from 8 to 43 GHz~\cite{Villata08}. The cause of these outbursts 
is unknown.

We shall also  apply {\tt RIDGE} to search for possible short-duration as well as long-duration 
gravitational wave signals from OJ 287 and S5 0716+71. For these objects,  since there were few observations per day, the on-source time interval 
is expected to be several days.

\section{Conclusions}
We described a method to search for  gravitational waves from blazars
and identified some sources which flared in TeV gamma rays and optical wavelengths, such as PKS 2155-304, OJ 287,  
and S5 0716+71. We discussed how a coherent network analysis algorithm ({\tt RIDGE}) can be used 
for this search. Sensitivity results with the {\tt RIDGE}  search algorithm, 
using simulated noise and simulated signals for the LIGO Hanford detectors 
and GEO detector, at the time of the PKS 2155-304 flare  are shown. 
This algorithm can detect $h_\mathrm{rss}$ of about $2.1 \times 10^{-22} Hz^{-1/2}$ for a detection 
probability of about 20 \%. This  corresponds to an energy estimate at the source of  $2.5 \times 10^{55}$ ergs
for the assumed signal waveform and should be of the same order of magnitude 
for any other short-duration burst signal. A complete end-to-end pipeline to  do this  search for gravitational waves from 
these blazar flares as well as other long-duration electromagnetic transients, which takes into 
account the motion  of the source across the sky has been designed.

\section*{Acknowledgments}
We would like to thank M~Eracleous, R~Frey, M~Hewitson, P~Kalmus,  
S~Marka, N~Leroy, L~Cadonati, P~Sutton, W~Cui, G~Ghishellini, K~Nilsson, and M~Valtonen  for  
many fruitful discussions and valuable  comments on this paper.
We are also indebted to W~Steffen 
for providing permission to show his animation in our poster at the GWDAW~12 workshop. This work is 
supported by NSF 428-51 29NV0 and Center for Gravitational Wave Physics (PSU), NSF PHY-055584 and NASA NAG5-13396 (UTB), 
NSF PHY-0653233 (SELU), Office of Scholarly Research (Andrews). The Center for Gravitational Wave Physics is funded by the National Science Foundation under Cooperative Agreement PHY 01-14375. This paper was assigned LIGO document number P080023.

\end{document}